\documentclass[12pt]{article}
\usepackage{epsf}

\newcommand{\suppressbynow}[1]{}
\begin{document}

\begin{titlepage}
\begin{flushleft}
       \hfill                      {\tt hep-th/9912209}\\
       \hfill                      UUITP-11/99\\
       \hfill                      HIP-1999-82/TH \\
       \hfill                      December 1999\\
\end{flushleft}
\vspace*{3mm}
\begin{center}
{\Large {\bf Black Hole Formation in AdS \\
and Thermalization on the Boundary \\}}
\vspace*{12mm}
{\large
Ulf H. Danielsson$^{ a,}$\footnote{E-mail: ulf@teorfys.uu.se} \\
Esko Keski-Vakkuri$^{ a,b,}$\footnote{E-mail: esko.keski-vakkuri@hip.fi\ \ \ 
{}$^b$ Permanent address.} \\
Mart\'{\i}n Kruczenski$^{ a,}$\footnote{E-mail: martink@teorfys.uu.se} \\ }

\vspace{5mm}

${}^a${\em Institutionen f\"{o}r teoretisk fysik \\
Box 803\\
S-751 08  Uppsala \\
Sweden \/}

\vspace{5mm}

${}^b${\em Helsinki Institute of Physics \\
P.O. Box 9\\
FIN-00014  University of Helsinki \\
Finland }
\vspace*{10mm}
\end{center}

%\maketitle

\begin{abstract}
We investigate black hole formation by a spherically collapsing thin
shell of matter in AdS space. This process has been suggested
to have a holographic
interpretation as thermalization of the CFT on the boundary of the AdS space.
The AdS/CFT duality relates the shell in the bulk to an off-equilibrium state 
of the boundary theory which evolves towards a thermal equilibrium when
the shell collapses to a black hole. We use 2-point
functions to obtain information about the spectrum of excitations
in the off-equilibrium state, and discuss how it characterizes the
approach towards thermal equilibrium. The full holographic
interpretation of the gravitational collapse would require a kinetic
theory of the CFT at strong coupling. We speculate that the kinetic
equations should be interpreted as a holographic dual of the equation
of motion of the collapsing shell. 
\end{abstract}

\end{titlepage}

\baselineskip16pt

\section{Introduction}

Following the discovery of the AdS/CFT correspondence \cite{Maldacena:1997re}
a large amount of work has been devoted to clarifying and extending its
implications. This work has been recently reviewed in \cite{Aharony:1999ti}
where an extensive list of references can be found. In this context, we have
investigated \cite{Keski-Vakkuri:1998nw, Danielsson:1998wt,
Danielsson:1999zt} the relation between physics in the bulk of AdS and its
holographic dual. We have mainly focused on the formulation in Minkowski
signature \cite{Balasub:1998sn,Banks:1998dd} since this allows the study of
dynamical processes. In particular we have been interested in the
holographic interpretation of black hole formation in the bulk since the
full understanding of this process is an important step towards the ultimate
goal of finding a holographic interpretation of the black hole information
problem \cite{Hawking:1976ra}, and (hopefully) its solution. We also believe
that by investigating the formation of black holes one will gain a deeper
understanding of general features of holography. The process that we are
focusing on is the spherical collapse of a thin shell of unspecified matter
in the bulk. A special case, the spherical 
collapse of a dust shell in AdS$_3$, was 
analyzed by Peleg and Steif in \cite{Peleg:1995wx}\footnote{The other
extreme case, collapse of a solid disk of dust in AdS$_3$ was investigated
by Mann and Ross in \cite{Ross:1993ba}.}. In the
holographic dual, a flat space field theory with no gravity, the bulk
process is expected to correspond to the time evolution and thermalization
of an initial out of equilibrium state \cite{ Banks:1998dd, Hawking:1983dh,
Balasub:1998de, Horowitz:1999gf}. The collapse of the shell due to gravity
describes the approach towards equilibrium, while the final formation of the
black hole corresponds to thermalization. It is intriguing how a
gravitational process in the bulk is encoded in a kinetic process in the
holographic dual. This, we believe, suggests a deep connection through
holography between gravity and kinetic theory of the boundary theory.

In \cite{Danielsson:1998wt} we developed Green function techniques that
allowed us to obtain the holographic ``images'' of various objects in the
bulk. In \cite{Danielsson:1999zt} we used these techniques to begin our
study of black hole formation. As a first step we studied the case of a very
large shell of matter which is just beginning to collapse. In this case we
could neglect the motion of the shell and use a quasistatic approximation.
The motivation for the study was to find out how the scale-radius duality 
\cite{Banks:1998dd, Balasub:1998de, Susskind:1998dq, Peet:1998wn} encodes
the radial size of a spherically symmetric shell, and to find properties of
the off-equlibrium state in the holographic dual. We found that the presence
of the shell manifested itself in the form of a series of poles in the
propagator. The location of these poles depends on the radius of the shell,
in agreement with the scale-radius duality. A slight surprise was that the
poles also had an imaginary part. Hence they can be interpreted as unstable
collective excitations, ``shellons'' (since they exist because of the
shell), in the off-equlibrium state.

In this paper we will carry the analysis one step further. While we are not
yet able to address the time-dependence of the collapse, there is another
regime where a quasistatic approach can again be adopted: the very last
stage when the shell is approaching the horizon radius $r_H$ of the black
hole about to form. There are two ways to justify the quasistatic
approximation.

First, in the final stage of the gravitational collapse the shell appears to
``freeze'' in the frame of an asymptotic observer. The approach of the
radius of the shell $r_{s}$ to the radius of the horizon $r_{H}$ is
exponentially slow, 
\begin{equation}
\frac{r_{s}-r_{H}}{r_{H}}\sim e^{-t/\tau _{H}}\ ,  \label{freeze}
\end{equation}
where the characteristic timescale $\tau _{H}$ is inversely proportional to
the Hawking temperature\footnote{%
This is true for the production of a generic black hole in a generic
dimension.}, 
\begin{equation}
\tau _{H}=\frac{1}{4\pi T_{H}}\ .
\end{equation}
Therefore, the quasistatic approximation is applicable for evaluating the
propagator as long as we focus on energies which are larger than the inverse
characteristic timescale, 
\begin{equation}
|\omega |>\frac{1}{\tau_H }\sim T_{H}\ .
\end{equation}
At the same time, this is the region of interest for investigating the
approach to thermal equilibrium. At equilibrium, the propagator becomes
thermal, with a characteristic infinite sequence of thermal poles 
\begin{equation}
\omega _{n}=i4\pi nT_{H}+\mathrm{const.}\ \ ,\ \ n=\pm 1,\pm 2,\ldots \ .
\end{equation}
When the radius of the shell is slightly larger than the horizon radius, the
boundary theory is slightly out of equilibrium, and one finds a correction
factor to the thermal propagator which has additional (shellon) poles and
zeroes. They characterize the deviation from the equilibrium; the poles and
zeroes flow together and disappear during the approach to equilibrium.

Second, one can avoid the restriction to large frequencies by investigating
a slightly different way of approaching the equilibrium. In the bulk one can
consider a sequence of static spherical shells with a constant total ADM
mass, but with progressively smaller radii approaching the horizon radius.
Instead of starting with a shell at rest and letting it collapse freely,
this corresponds to slowing down the collapse by applying an external force
so that the process becomes quasistatic, while increasing the amount of
matter in the shell. The ADM energy needs to be kept fixed to ensure that
the shell approaches its horizon radius, therefore the loss of kinetic
energy needs to be compensated by adding matter to the shell during the
process. In this case one obtains more detailed information about the
propagator in the boundary theory. Presumably, the quasistatic approximation
amounts in the boundary theory to replace the kinetic theory by a
thermodynamical treatment of small deviations from equilibrium. In
thermodynamics the evolution of a system is always considered to be
quasistatic, and small deviations from equilibrium can be arranged to be
such by applying external forces. The evolution is then determined by the
gradient of entropy depending on a small parameter characterizing the
deviation from equilibrium, which should correspond to the applied force
balancing the gravitational force in the bulk. In contrast, recently \cite
{deBoer:1999} the equations of motion in the bulk were related to
renormalization group equations in the boundary. This is a good
interpretation for the static background geometry but general relativity
also determines the time evolution of an object in the bulk -- this is a
dynamical process related to the dynamics of the boundary as discussed above.

The paper is organized as follows. In section 2 we review our previous work
and compare our results with other recent work on black hole formation \cite
{Balasubramanian:1999zv, Giddings:1999zu, Horowitz:1999jd, Horowitz:1999uv}.
The analysis of the last stages of black hole formation in the case of AdS$%
_{3}$ is given in section 3. Section 4 contains a discussion of results and
some ideas for the future.

\section{Derivation of the correlators}

We begin by briefly reviewing (following \cite{Gubser:1998bc,Witten:1998qj}
and as described in \cite{Danielsson:1999zt}) the essential steps of the
computation of a two point function of a boundary operator coupling to a
bulk scalar field in an asymptotically AdS spacetime. A scalar field in the
bulk satisfies the classical equation of motion 
\begin{equation}
\left[ \frac{1}{\sqrt{-g}}\partial _{\mu }\sqrt{-g}g^{\mu \nu }\partial
_{\nu }-m^{2}\right] \ \phi (t,\vec{x},r)=0  \label{fieldeqn}
\end{equation}
where $g_{\mu \nu }$ is the bulk metric and the mass term $m^{2}$ contains
the mass of the field, the coupling to the curvature scalar\footnote{%
When we consider \emph{e.g.} a spacetime of a collapsing shell in AdS$_{d+1}$
with $d>2$, the exterior metric has a nonconstant scalar curvature. In such
cases we assume that the scalar field is minimally coupled so that $m^{2}$
remains a constant.}, and contributions from a Kaluza-Klein reduction on a
compact internal space, if the original spacetime is higher dimensional. At
first, we choose the desired coordinate system for (a part of) AdS space. We
will work in Minkowski signature. Depending on the choice of coordinates, we
then impose the appropriate boundary condition for the bulk field in the
interior. For example, in AdS with global coordinates one requires that $%
\phi $ vanishes at the origin. In Poincar\'{e} coordinates or in the case of
a black hole, the appropiate boundary condition at the horizon is an ingoing
wave when the imaginary part of the energy is positive and an outgoing when
it is negative. The propagator then has a cut on the real axis where the
imaginary part changes sign. Let $t,\vec{x}$ denote the coordinates that
parametrize the boundary, and let $r$ denote the radial coordinate. We
perform a Fourier transformation to momentum space in the boundary
coordinates, so the interior solution is $\phi (\omega ,\vec{k},r)$. This
solution is uniquely determined (up to an overall normalization) by the
boundary conditions discussed above.

We then study its behavior near the boundary. In Minkowski signature, there
are two kinds of bulk field modes, normalizable and non-normalizable%
\footnote{%
The Minkowski signature version of AdS/CFT duality was first investigated in 
\cite{Balasub:1998sn,Banks:1998dd}.}. In momentum space, we denote the
former by $\phi^{(+)}(\omega ,\vec{k},r)$ and the latter by $\phi
^{(-)}(\omega ,\vec{k},r)$. We normalize the modes such that their
asymptotic behavior near the boundary is 
\begin{equation}
\phi ^{(\pm )}(\omega ,\vec{k},r)=r^{-\Delta _{\pm }}\cdot 1\ \
(r\rightarrow \infty )
\end{equation}
where 
\begin{equation}
\Delta _{\pm }=\frac{1}{2}(d\pm \sqrt{d^{2}+4m^{2}})\equiv \frac{d}{2}\pm
\nu \ .  \label{eq:Delta}
\end{equation}
Now, in the asymptotic region, the interior solution $\phi (\omega ,\vec{k}%
,r)$ becomes a linear combination of the normalizable and non-normalizable
solutions, with an asymptotic behavior 
\begin{equation}
\phi (t,\vec{x},r)\approx r^{-\Delta _{+}}C(\omega ,\vec{k})+r^{-\Delta
_{-}}D(\omega ,\vec{k})\ .  \label{asy}
\end{equation}
The coefficient $D(\omega ,\vec{k})$ is the Fourier transform of the
boundary data $\phi _{0}(t,\vec{x})$ which acts as a source term for a
dimension $\Delta _{+}$ operator $\mathcal{O}(t,\vec{x})$ in the boundary
theory \cite{Gubser:1998bc, Witten:1998qj}. Next, we rewrite (\ref{asy}) in
the form 
\begin{equation}
\phi (\omega ,\vec{k},r)\approx \left[ r^{-\Delta _{+}}G(\omega ,\vec{k}%
)+r^{-\Delta _{-}}\right] \phi _{0}(\omega ,\vec{k})\ ,
\end{equation}
where 
\begin{equation}
G(\omega ,\vec{k})=\frac{C(\omega ,\vec{k})}{D(\omega ,\vec{k})}\ .
\label{gee}
\end{equation}
One can then check that the (finite part) of the two point function of the
operator $\mathcal{O}$ at the boundary has the momentum space form\footnote{%
The evaluation of the overall coefficient involves a subtlety, see \cite
{Freedman:1998tz}. In the remainder of the paper we will suppress overall
coefficients as they do not affect the pole structure of the propagator.} 
\begin{equation}
\langle \mathcal{O}(\omega ,\vec{k})\mathcal{O}(\omega ^{\prime },\vec{k}%
^{\prime })\rangle =\left( \frac{\Delta _{+}-\Delta _{-}}{2}\right) \delta
(\omega +\omega ^{\prime })\delta (\vec{k}+\vec{k}^{\prime })G(\omega ,\vec{k%
})\ .
\end{equation}

In the case that the geometry has a horizon, the propagator, as a function
of the frequency $\omega $, has a cut on the real axis which arises from the
boundary conditions as we already discussed. In that case one can
analytically continue the propagator from the upper half-plane into the
lower half-plane. This analytic continuation can have poles on the lower
half plane signaling the presence of unstable quasi particles.

Next, we compute the propagator in the presence of a thin matter shell of
radius $r_{s}$ in AdS$_{d+1}$ space. We are going to restrict attention to a
quasistatic approximation: we assume that the shell is collapsing very
slowly (as measured in the asymptotic time), so that it appears to be static%
\footnote{%
Of course, a complete treatment should take into account what happens when
the shell is collapsing rapidly. This can be expected to give rise to
particle production in the boundary theory even if we neglect the particle
production in the bulk. We will leave these more demanding issues aside here.%
}. The static case has an interest also of its own since it provides a
series of states which are closer and closer to equilibrium. Understanding
how these states differ from the black hole from the boundary point of view,
gives an insight into the boundary processes at work. Furthermore, the usual
thermodynamical approach to out of equilibrium states is to consider states
which are at equilibrium but for a variable that is externally kept fixed.
In this case the variable is the radius of the shell.

In the interior of the shell, the metric is the AdS metric in global
coordinates, and the exterior metric is the AdS black hole metric. Both
metrics can be expressed in the form 
\begin{equation}
ds^{2}=-f(r)dt^{2}+\frac{dr^{2}}{f(r)}+r^{2}d\Omega _{d-1}\ ,  \label{metric}
\end{equation}
with the function $f(r)$: 
\begin{equation}
f(r)=\left\{ 
\begin{array}{lcl}
f_{1}(r)=1+r^{2} & \mathrm{if} & r<r_{s} \\ 
f_{2}(r)=1-\frac{\mu }{r^{d-2}}+r^{2} & \mathrm{if} & r>r_{s}\ ,
\end{array}
\right.
\end{equation}
where the parameter $\mu $ is related to the ADM mass of the black hole. 
The radius of the horizon of the black hole, $r_{H},$ is determined by
solving the equation $f_{2}(r_{H})=0$. (We are using units where $t,r$ and
the angular coordinates are dimensionless and the AdS radius is set to $R=1$%
. The dimensions can be restored by replacing everywhere $r\rightarrow r/R$
etc. With the metric (\ref{metric}), the scalar field equation (\ref
{fieldeqn}) can be reduced (after a Fourier transformation to momentum space
in the boundary coordinates) to the radial equation 
\begin{equation}
\frac{1}{r^{d-1}}\partial _{r}\left( r^{d-1}f(r)\partial _{r}\right) \phi
(\omega ,\vec{k},r)+\left( \frac{\omega ^{2}}{f(r)}-\frac{k^{2}}{r^{2}}%
-m^{2}\right) \phi (\omega ,\vec{k},r)=0\ ,  \label{radeqn}
\end{equation}
where $k^{2}=l(l+d-2),\ l=0,1,2,\ldots $ is the eigenvalue of the Laplacian
on the $(d-1)$-sphere. The interior and exterior solutions $\phi _{1}$ and $%
\phi _{2}$ of the radial equation must be matched across the thin shell
using flux conservation. The matching conditions are 
\begin{eqnarray}
\phi _{1}|_{r=r_{s}} &=&\phi _{2}|_{r=r_{s}} \\
f_{1}(r)\partial _{r}\phi _{1}|_{r=r_{s}} &=&f_{2}(r)\partial _{r}\phi
_{2}|_{r=r_{s}}\ .
\end{eqnarray}
To find the propagator $G(\omega ,\vec{k})$ in the boundary, we first expand
the exterior solution in normalizable and non-normalizable modes, 
\begin{equation}
\phi _{2}(\omega ,\vec{k},r)\approx r^{-\Delta _{+}}C(\omega ,\vec{x}%
)+r^{-\Delta _{-}}D(\omega ,\vec{k})\ ,  \label{asy2}
\end{equation}
then substitute (\ref{asy2}) in the matching conditions above, solve for the
ratio $C/D$ in terms of $\phi _{1}$ and $\phi _{2}$, and substitute into
equation (\ref{gee}) which gives the propagator. The result is 
\begin{equation}
G(\omega ,\vec{k})=-\frac{\phi _{1}f_{2}\partial _{r}\phi _{2}^{(-)}-\phi
_{2}^{(-)}f_{1}\partial _{r}\phi _{1}}{\phi _{1}f_{2}\partial _{r}\phi
_{2}^{(+)}-\phi _{2}^{(+)}f_{1}\partial _{r}\phi _{1}}\ .  \label{propag}
\end{equation}
In the above, all quantities are evaluated at the radius of the shell $r_{s}$%
. The analytic structure of the resulting propagator gives information about
the spectrum of excitations in the boundary theory. In particular, we are
interested in the poles of the propagator at zero transverse momentum ($\vec{%
k}=0$) in the complex $\omega $-plane. Poles at positive real $\omega $-axis
correspond to masses of stable composite objects created and annihilated by $%
\mathcal{O}$. However, we will typically find complex poles at 
\begin{equation}
\omega _{n}=M_{n}-\frac{i}{2}\Gamma _{n}
\end{equation}
corresponding to unstable resonances created and annihilated by $\mathcal{O}$%
. In a previous paper\cite{Danielsson:1999zt}, we used the name ``shellons''
for the resonances. The real part $M_{n}$ gives the mass of the resonance,
and the imaginary part $\Gamma _{n}$ is the width which is the inverse of
the lifetime. There can also be poles in the imaginary $\omega $-axis, for
example when the black hole has fully formed and the boundary theory has
thermalized. Then the boundary propagator becomes a thermal propagator, and
the imaginary poles are related to periodicity in imaginary time.

{}From (\ref{propag}), we see that the poles typically correspond to the
zeroes of the denominator (unless they cancel against those of the
numerator). The vanishing of the denominator means that the interior
solution $\phi _{1}$ matches completely with a normalizable exterior
solution $\phi _{2}^{(+)}$. Thus, the total mode is normalizable -- unless $%
\omega $ is complex, in which case it will be quasinormalizable. For a
better understanding of the situation, it is instructive to map the radial
equation into the form of a Schr\"{o}dinger equation and consider the
resulting quantum mechanical analogue. We rescale the field $\phi $ as 
\begin{equation}
\phi (\omega ,\vec{k},r)=r^{(-d+1)/2}\psi (\omega ,\vec{k},r)
\end{equation}
and use a ``tortoise'' coordinate 
\begin{equation}
r_{\ast }=\int \frac{dr}{f(r)}\ .  \label{tort}
\end{equation}
Note that we have to use $f_{1}$ in the interior and $f_{2}$ in the
exterior, so we must add an integration constant to ensure that $r_{\ast }$
is continuous across the shell. The range of $r_{\ast }$ is finite. For
example, if $d=2$, we get 
\begin{equation}
r_{\ast }=\left\{ 
\begin{array}{ll}
-\frac{1}{2r_{H}}\ln \left( \frac{r+r_{H}}{r-r_{H}}\right) & \ \ \ \ r>r_{s}
\\ 
\arctan (r)-r_{0} & \ \ \ \ 0\leq r<r_{s}
\end{array}
\right.
\end{equation}
where 
\begin{equation}
r_{0}=\arctan (r_{s})+\frac{1}{2r_{H}}\ln \left( \frac{r_{s}+r_{H}}{%
r_{s}-r_{H}}\right)
\end{equation}
and the full range of $r_{\ast }$ is the finite interval $-r_{0}\leq r_{\ast
}\leq 0$. In the following we will mostly be considering the case where the
radius of the final black hole, $r_{H}$, and therefore also the radius of
the shell, $r_{s}$, are very large compared to the AdS radius ($r_s>r_H\gg1$
in our units). The radial equation in the interior will then reduce to that
in a Poincar\'{e} patch of AdS space (see \cite{Danielsson:1999zt}), and the
two equations above become 
\begin{equation}
r_{\ast }=\left\{ 
\begin{array}{c}
\begin{array}{ll}
-\frac{1}{2r_{H}}\ln \left( \frac{r+r_{H}}{r-r_{H}}\right) & \ \ \ \ r>r_{s}
\end{array}
\\ 
\begin{array}{ll}
-\frac{1}{r}-r_{0} & \ \ \ \ 0\leq r<r_{s}
\end{array}
\end{array}
\right.
\end{equation}
where 
\begin{equation}
r_{0}=\frac{1}{2r_{H}}\ln \left( \frac{r_{s}+r_{H}}{r_{s}-r_{H}}\right) -%
\frac{1}{r_{s}}\ .
\end{equation}
The full range of $r_{\ast }$ is now the infinite interval $-\infty \leq
r_{\ast }\leq 0$, corresponding to the infinite AdS throat.

With the above rescalings, the radial equation (\ref{radeqn}) reduces to the
form of a time independent Schr\"{o}dinger equation, 
\begin{equation}
-\partial _{r_{\ast }}^{2}\psi +V(r_{\ast })\psi =\omega ^{2}\psi
\end{equation}
with a potential 
\begin{equation}
V(r_{\ast })=\frac{(d-1)(d-3)f^{2}}{4r^{2}}+\frac{(d-1)\partial _{r_{\ast }}f%
}{2r}+\frac{(k^{2}+m^{2}r)f}{r}\ ,
\end{equation}
where $r=r(r_{\ast })$ from inverting (\ref{tort}) and $f=f(r(r_{\ast }))$.
Thus the problem of solving the radial equation becomes analogous to a one
dimensional quantum mechanical scattering problem. In Figure 1 a typical
profile of the 
\begin{figure}[tbp]
\epsfysize=8.0truecm \epsffile{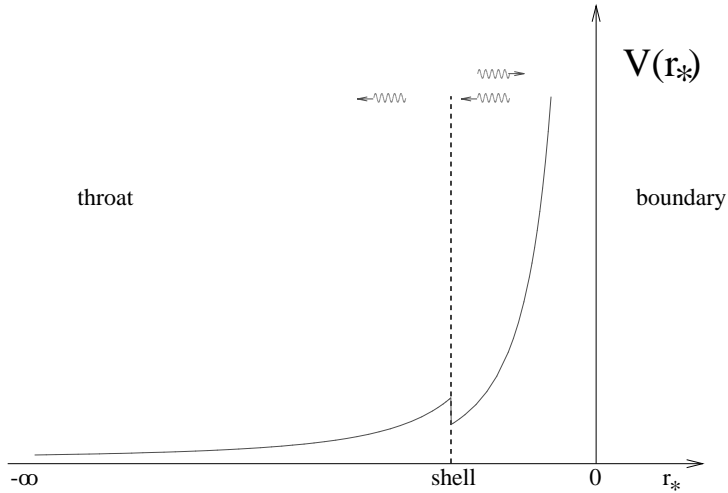}
\caption{Potential in the presence of a shell.}
\label{fig:potential}
\end{figure}
potential is depicted. It blows up at the boundary and has a small barrier
at the location of the shell\footnote{%
If we were relax the quasistatic approximation and allow for the time
dependence of the collapsing shell, we would need to consider quantum
mechanical scattering with a time dependent potential: the barrier would be
moving to the left.}. There is also an infinite well at $r_*=-r_0$, but when
we consider a large black hole the well is pushed down the AdS throat
towards $r_*=-\infty $ and plays no role in the calculation. Now let us
return to the poles of (\ref{propag}). As we discussed above, the
appropriate interior boundary condition for $\phi _{1}$ is that it is a
purely ingoing wave. On the other hand, the normalizable mode in the
exterior is a superposition of an outgoing wave and an ingoing wave, due to
the reflection from the boundary of AdS. For example, it is simple to check
this by looking at the high frequency limit of the normalizable mode $\phi
_{2}^{(+)}$ \cite{Keski-Vakkuri:1998nw, Danielsson:1999zt}, which looks like 
\begin{equation}
\phi _{2}\approx r^{(-d+1)/2}\cos (\frac{\omega }{r}+\theta _{0})
\end{equation}
where $\theta _{0}$ is a constant phase shift. For the ``wavefunction'' $%
\psi $ this means that it is a superposition of ingoing and outgoing waves
with \emph{equal amplitude} on the right hand side of the barrier
corresponding to the shell (see Figure 1) and on the left hand side there is
a purely ingoing wave. Thus there is total reflection together with
transmission -- it is impossible to satisfy both conditions with real energy
levels. However, it is possible with \emph{complex} energy levels $\omega
^{2}$. This is another way to see why the poles appear at complex
frequencies $\omega $. Incidentally, note that using the tortoise coordinate
(\ref{tort}), the propagator (\ref{propag}) takes an even simpler form - it
is just a ratio of two Wronskians 
\begin{equation}
G(\omega ,\vec{k})=-\frac{W(\phi _{1},\phi _{2}^{(-)})}{W(\phi _{1},\phi
_{2}^{(+)})}
\end{equation}
evaluated at $r_{\ast }=r_{\ast }(r_{s})$.

Let us compare our approach with other recent work. Horowitz and Hubeny \cite
{Horowitz:1999jd} investigated the quasinormal modes in the background of a
black hole\footnote{%
Related work can be found in \cite{KalyanaRama:1999zj}.}. They study the
equation (\ref{radeqn}) with the black hole metric ($f(r)=f_{2}(r)$). The
boundary conditions in \cite{Horowitz:1999jd} for a scalar field are very
similar to ours: one takes the field to be a superposition of outgoing and
ingoing modes near the boundary (normalizable modes), whereas the interior
boundary condition in \cite{Horowitz:1999jd} is that there should be only
ingoing modes near the horizon of the black hole. In our case with a shell,
the interior metric is that of a pure AdS space. However, if the radius of
the shell is larger than the AdS space ($r_{s}\gg 1$ in the present units),
a Poincar\'{e} horizon appears at $r\rightarrow 0$ ($r_{\ast }=-\infty $)
due to the approximation, and there is a natural choice between ingoing or
outgoing modes, which is reflected in a cut appearing in the propagator%
\footnote{%
The choice between the two sides of the cut corresponds to a choice between
a retarded or advanced propagator.} of the boundary operator $\mathcal{O}$.
Thus, on a formal level the modes we discussed in \cite{Danielsson:1999zt}
are analogous to the quasinormal modes of black holes \cite{Horowitz:1999jd}%
. The complex spectrum arises because of the choice of boundary conditions,
and the main difference between \cite{Horowitz:1999jd} and \cite
{Danielsson:1999zt} is that the horizon of a black hole has been replaced by
a Poincar\'{e} horizon. However, the physical interpretation appears to be
somewhat different. The horizon of a black hole is a real boundary through
which matter can fall in and be lost, while the appearance of a Poincar\'{e}
horizon in the above is merely an artifact of the approximation since we
started from global AdS space. In reality we would expect an ingoing wave to
be reflected back from the origin of the AdS space and not to be lost. What
happens is that if the radius of the shell is much larger than the AdS
radius, $r_{s}\gg 1$, and the frequency is complex (with $\Im\omega<0$) the
reflection is exponentially attenuated, so that it can be suppressed. There
are two other ways to arrange for the ingoing mode from the shell not to
come back at all. One way is simply to consider a shell falling into an
existing (small) black hole instead of creating the black hole in the
collapse. Another way is to consider late times in the collapse when the
shell is approaching the radius of the horizon of the black hole about to be
created. From the space-time diagram of black hole formation in a spherical
collapse, one can see that the horizon forms even before the black hole has
formed. Therefore, the ingoing modes from the shell will be lost into the
region of trapped surfaces and never come back.

Another related work is by Giddings and Ross \cite{Giddings:1999zu} who
investigated a collapsing shell of D-branes. In their case, contrary to ours
with an unspecified matter shell in AdS space, the geometry in the interior
of the D-brane shell is flat. Hence they do not have an infinite throat,
where objects falling in could get lost. Instead, their space is cut off at
a finite $r_{\ast }$. As a consequence, when they investigate the poles of
the propagator, they initially obtain a discrete spectrum of stable
excitations with real energies, much like the global modes in our case.
However, the shell itself can absorb incoming waves, and after considering
this effect they also obtain imaginary contributions to the energy spectrum.
So the absorption by the shell plays a similar role as the absorption by the
infinite throat in AdS space in our case.

Furthermore, Balasubramanian and Ross \cite{Balasubramanian:1999zv} have
investigated black hole formation in AdS$_3$ by point particles. The bulk
solution was investigated by Matschull \cite{Matschull:1998rv}, and
Balasubramanian and Ross investigate the use of 2-point functions in the
boundary theory to keep track of the positions of the point particles as
they approach each other. In this case one does not have the complication of
spherical symmetry, the point particle positions can be read off from kinks
in the 2-point functions in \emph{coordinate} space.

Finally, we would like to mention a slightly different line of work on
thermalization in the context of black holes and AdS/CFT correspondence. 
Kiritsis and Taylor \cite{Kiritsis:1999ke} investigated D-brane probes falling
into a black hole. In the process, the potential energy of the probe is
converted into heat, which then is absorbed by the black hole. By investigating
the bulk action of the probe, one can {\em e.g.} derive the equation of
state for the black hole. But, on the other hand, one can study the
dual gauge theory effective action of the probe at finite temperature and
try to reproduce predictions from bulk calculations; this was investigated
in \cite{Kiritsis:1999tx}.

\section{Black hole formation in AdS$_{3}$}

In this section we examine the boundary propagator in the case that a shell
is quasistatically forming a black hole in AdS$_{3}$ space. As in \cite
{Danielsson:1999zt}, we assume that the resulting black hole has a much
larger radius than that of the AdS space: $r_{H}\gg 1$ in the present units.
The problem is to find the modes $\phi _{1},\phi _{2}^{(\pm )}$ which are
needed in the propagator formula (\ref{propag}). To find the interior mode,
note that in the region $r>r_{s}\gg 1$ we can approximate $%
f_{1}(r)=1+r^{2}\approx r^{2}$. Then the radial equation becomes the mode
equation in Poincar\'{e} coordinates, and the mode solutions will be Bessel
functions $\phi \sim (1/r)J_{\pm \nu }(\sqrt{\omega ^{2}-k^{2}}/r)$. As
discussed in section 2, in Poincar\'{e} coordinates the natural choice for
the interior boundary condition is between ingoing and outgoing modes, and
we choose the former, as we did in \cite{Danielsson:1999zt}. However, in 
\cite{Danielsson:1999zt} we made an additional approximation and focused on
the large frequency limit $\omega /r\gg 1$. Here we will not make that
approximation, so the infalling mode is the following linear combination\ 
\begin{equation}
\phi _{1}=c_{1}\frac{1}{r}(J_{\nu }\left( \frac{\sqrt{\omega ^{2}-k^{2}}}{r}%
\right) -e^{i\pi \nu }J_{-\nu }\left( \frac{\sqrt{\omega ^{2}-k^{2}}}{r}%
\right) )\ ,  \label{phi1}
\end{equation}
where $\nu =\sqrt{1+m^{2}}$ and $c_{1}$ is a normalization constant (which
will drop out in the answer for the propagator).

Then, we need the exterior modes $\phi _{2}^{(\pm )}$ in BTZ coordinates. In
our previous paper \cite{Danielsson:1999zt}, we used a WKB approximation in
the large frequency limit to obtain the modes. Here, we will instead use the
exact solutions of the wave equation \cite{Ichinose:1995rg}, the
normalizable and nonnormalizable modes can be found in \cite
{Keski-Vakkuri:1998nw}. If the shell is large ($r_{s}\gg r_{0}$), we use 
\begin{equation}
\phi _{2}^{(\pm )}=(u-1)^{\alpha }u^{-(\Delta _{\pm }/2)-\alpha }F(\alpha + 
\frac{\Delta _{\pm }}{2},\alpha +\frac{\Delta _{\pm }}{2};1+\nu ;u^{-1})\ ,
\label{phi2large}
\end{equation}
where $u=r^{2}/r_{H}^{2}$, $\alpha=i\omega/(2r_H)$ and $\Delta_\pm$ were
defined in (\ref{eq:Delta}). Substituting (\ref{phi1}) and (\ref{phi2large})
into (\ref{propag}), we obtain the result for the propagator of the operator 
$\mathcal{O}$ on the boundary in the presence of a large matter shell in the
bulk. Using the latter result we checked that we reobtain our previous
approximate result for the poles in the $r_{s}\gg r_{H},\ \omega \gg r_{s}$
limit: 
\begin{equation}
\omega _{n}=\pi r_{s}(n+\frac{3}{4}+\frac{\nu }{2})-\frac{ir_{s}}{2}\ln
\left( \frac{4\pi nr_{s}^{2}}{r_{H}^{2}}\right) \ .  \label{shellons}
\end{equation}
This is sufficient to illustrate the main point: the presence of the
spherically symmetric shell in the bulk is encoded in a tower of resonances
in the boundary theory, with the above spectrum from which one can read
their masses and decay widths. 
An intriguing thing about the shellon poles 
becomes apparent if we focus on the UV limit $|\omega|\gg 1$. By the
scale-radius duality one might have thought that we in the bulk are probing
only the geometry very near the boundary. But that is a region outside of
the shell, so the geometry is equal to the black hole geometry. On the other
hand, near the boundary the black hole metric reduces to the same form as
the Poincar\'{e} metric. From these arguments we might have expected that
the analytic structure of the propagator in the UV limit is either similar
to that of the thermal propagator, or the Poincar\'{e} propagator\footnote{%
We would like to thank Per Kraus and Sandip Trivedi for pointing out this
issue to us.}. On the other hand, from (\ref{shellons}) we know that
there are poles for arbitrary large $\omega$ arising from the shell. 
A possible interpretation is that the poles of the 
propagator do not only carry local information about the geometry of 
spacetime near the boundary - they also carry information about an event 
that happens deep in the bulk inside the shell: the absorption of the 
ingoing wave into the (fictitious) horizon, which was used as a 
boundary condition in the evaluation of the propagator.

Rather than present more details and more
accurate results for the poles in the large shell case, we turn our
attention to the other extreme of the collapse of the shell: the approach to
the horizon radius ($(r-r_{H})/r_{H}\ll 1$) where we expect to see
thermalization on the boundary.

As the shell is approaching the horizon, we choose the following expressions
for the exterior normalizable and nonnormalizable modes \cite
{Keski-Vakkuri:1998nw}: 
\begin{eqnarray}
\phi^{(\pm)}_2 &=& A_{\pm} (u-1)^{\alpha} F(\alpha+\frac{\Delta_{\pm}}{2}%
,\alpha+\frac{\Delta_{\mp}}{2}; 2\alpha +1;1-u)  \nonumber \\
\mbox{} &+& B_{\pm} (u-1)^{-\alpha} F(-\alpha+\frac{\Delta_{\pm}}{2},-\alpha+%
\frac{\Delta_{\mp}}{2}; -2\alpha +1;1-u) \ ,  \label{phi2near}
\end{eqnarray}
where 
\begin{eqnarray}
A_{\pm} &=& \frac{\Gamma (1\pm \nu)\Gamma(-2\alpha )} {\Gamma^2 (-\alpha +%
\frac{\Delta_{\pm}}{2})} \\
B_{\pm} &=& \frac{\Gamma (1\pm \nu)\Gamma(2\alpha )} {\Gamma^2 (\alpha +%
\frac{\Delta_{\pm}}{2})} \ .
\end{eqnarray}
and we used $\alpha$ as defined in (\ref{phi2large}). We then plug the
interior mode (\ref{phi1}) and the exterior modes (\ref{phi2near}) into the
main propagator formula (\ref{propag}), and study its analytic structure. To
simplify matters, we again focus on the propagator at zero transverse
momentum, $G(\omega ,k=0)$.

Let us begin with some general observations. As the shell approaches the
horizon, the boundary theory gets closer and closer to thermal equilibrium.
Therefore, in the limit $r_{s}\rightarrow r_{H}$, $G(\omega ,k)$ should
become the thermal propagator which we evaluated in our previous work \cite
{Danielsson:1999zt}. The thermal propagator (which we denote by $%
G_{T}(\omega ,k)$ from now on) has a characteristic infinite sequence of
zeroes and poles, at zero momentum they are located on the imaginary axis on
the complex $\omega $-plane. Before the limit $r_{s}\rightarrow r_{H} $,
when the shell is at a small distance from the horizon, the boundary theory
is slighty off thermal equilibrium. In that case, we find it useful to write
the propagator in a factorized form 
\begin{equation}
G(\omega ,k=0)=G_{T}(\omega ,k=0)\times H(\omega ,k=0)\ ,
\end{equation}
This is useful for two reasons. First, if $r_{s}\rightarrow r_{H}$ and $%
\mathrm{Im} \omega >0$, then $H(\omega ,k=0)\rightarrow 1$ and $H$ encodes
small deviations from the thermal equilibrium when $r_{s}\neq r_{H}$.
Second, when $\mathrm{Im} \omega <0$, $H(\omega ,k=0)$ has poles which are
then poles of the propagator. The expression can be misleading if used on,
for example, the imaginary axis, since $G_{T}$ has poles at $\omega =-i4\pi
n T_H$ ($T_H$ is the temperature of the black hole) whereas $G$ does not
since they are canceled by $H$. These thermal poles appear in $G$ only after
taking the limit $r_{s}\rightarrow r_{H}$. Formally one has to consider $%
\mathrm{Im} \omega >0$, take the limit $r_{s}\rightarrow r_{H}$ in which
case $H(\omega ,k=0)\rightarrow 1 $, and then analytically continue below
the real axis. Since then $G=G_{T}$ they also have the same poles. The
analytic continuation is necessary since when $r_{s}\rightarrow r_{H}$ a new
cut arises as an accumulation of poles and zeros. In fact it can be said
that the cut due to the boundary conditions at the Poincar\'{e} horizon is
replaced by another one due to the new boundary conditions at the horizon of
the black hole which forms.

After discussing this point, we try to analyze the propagator without any
approximations for the interior and exterior modes (\ref{phi1}) and (\ref
{phi2near}). We assume that we are initially on the upper half $\omega $%
-plane. Then, one can see that the leading terms of the exterior modes (\ref
{phi2near}) have a coefficient $A_{\pm }$. We can pull out these
coefficients in the numerator and denominator of $G$, whereafter they will
give us the ``thermal'' factor $G_{T}$. In other words, now 
\begin{equation}
G_{T}(\omega ,0)=-\frac{A_{-}}{A_{+}}=-\left( \frac{\Gamma (-\frac{i\omega }{%
4\pi T}+\frac{\Delta _{+}}{2})}{\Gamma (-\frac{i\omega }{4\pi T}+\frac{%
\Delta _{-}}{2})}\right) ^{2}
\end{equation}
and 
\begin{equation}
H(\omega ,0)=\frac{\phi _{1}f_{2}\partial _{r}(\phi _{2}^{(-)}/A_{-})-(\phi
_{2}^{(-)}/A_{-})f_{1}\partial _{r}\phi _{1}}{\phi _{1}f_{2}\partial
_{r}(\phi _{2}^{(+)}/A_{+})-(\phi _{2}^{(+)}/A_{+})f_{1}\partial _{r}\phi
_{1}}\ .
\end{equation}
The ``correction factor'' $H(\omega ,0)$ is too complicated to allow us to
find an analytic expression for the poles. Therefore we have evaluated them
numerically, using Maple V version 5. Figure 2 below depicts the flow of the
first couple of poles of $H$ as the shell radius $r_{s}$ approaches the
horizon radius $r_{H}$. 
\begin{figure}[tbp]
\epsfysize=7.0truecm \epsffile{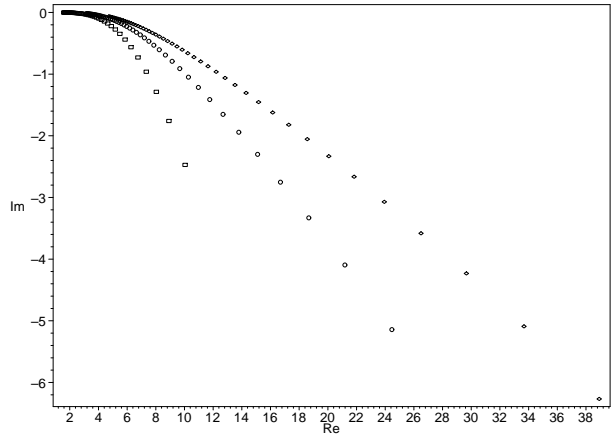}
\caption{Flow of the first three poles of H in the complex $\protect\omega$
plane as the shell contracts towards the horizon. In the numerical
calculation, we used the parameter values $\protect\nu=1.5$, $r_H=10$
(horizon radius), and the initial radius of the shell is $r_s=10.25$.}
\label{fig:sketch}
\end{figure}
Numerical investigations tell us that the poles of $H$ start out in the
lower half complex $\omega $-plane (where the poles were found as the shell
was still large), and then move towards the real axis and the origin.
Specifically, as $r_{s}$ is sufficiently close to $r_{0}$, the real and
imaginary parts of the poles $\omega $ become much smaller than $r_{H},r_{s}$%
. Further, the imaginary part decreases more rapidly than the real part, and
quickly becomes many orders of magnitude smaller. This is illustrated in
Table 1 below which lists a few characteristic values for a pole as $r_s$
approaches $r_H$.

\begin{center}
\begin{tabular}{|l|l|}
\hline
$r_s$ & $\omega$ \\ \hline\hline
10.25 & 10.04-2.47i \\ \hline
10.016 & 6.76-0.73i \\ \hline
10.0020 & 5.49-0.35i \\ \hline
10.00024 & 4.65-0.18i \\ \hline
10.000031 & 4.03-0.10i \\ \hline
10.0000020 & 3.43-0.049i \\ \hline
10.00000024 & 3.08-0.030i \\ \hline
10.000000030 & 2.80-0.019i \\ \hline
\end{tabular}
\vskip 0.6cm \textbf{Table 1.} Pole values $\omega$ as $r_s$ approaches $%
r_H=10$. (Using $\nu=1.5$.)
\end{center}

We now turn to search for a simplified expression for $H$, and begin by
making some approximations for the interior and exterior modes. Let's first
consider the exterior modes $\phi _{2}^{(\pm )}$. As $(r_{s}-r_{H})/r_{H}$
becomes small, $u=r_s^2/r_H^2$ approaches $1$ and we can do a series
expansion in $(u-1)$ obtaining 
\begin{equation}
\phi _{2}^{(\pm )}\approx \left\{ A_{\pm } \left[(u-1)^{\alpha }+\mathcal{O}%
((u-1)^{\alpha +1})\right]+ B_{\pm }\left[(u-1)^{-\alpha }+\mathcal{O}%
((u-1)^{-\alpha +1})\right]\right\} \ .  \label{p2app1}
\end{equation}
Similarly, we obtain a series expansion for the combination $%
f_{2}\partial_{r}\phi _{2}^{(\pm )}$ which also appears in the equation (\ref
{propag}) for the propagator, 
\begin{equation}
f_{2}\partial _{r}\phi _{2}^{(\pm )}\approx 2r\left\{A_{\pm }\left[
(u-1)^{\alpha }+\mathcal{O}((u-1)^{\alpha +1})\right]-B_{\pm }\left[%
(u-1)^{-\alpha }+\mathcal{O}((u-1)^{-\alpha +1})\right]\right\} \ .
\label{p2app2}
\end{equation}
Recalling that $\alpha=i\omega/(2r_H)$ it immediately follows that $%
(u-1)^{\alpha }$ is the leading term if $\mathrm{Im} \omega >0$ whereas $%
(u-1)^{-\alpha }$ dominates for $\mathrm{Im} \omega <0$. Replacing in
equation (\ref{propag}) results in 
\begin{equation}
\begin{array}{lcl}
\lim_{r_s\rightarrow r_H} G(\omega,r_s) & = & \left\{ 
\begin{array}{lcl}
-\frac{A_-}{A_+} & \mbox{if} & \mathrm{Im} \omega >0 \\ 
-\frac{B_-}{B_+} & \mbox{if} & \mathrm{Im} \omega <0
\end{array}
\right.
\end{array}
\end{equation}
Note that independently of the function $\phi_1$, i.e. independently of the
boundary conditions at the Poincar\'{e} throat, the propagator develops a
cut with the right boundary conditions at the horizon of the black hole. At
the cut the imaginary part of the propagator changes sign since $%
A_\pm=B_\pm^*$.

After checking that in the limit $r_s\rightarrow r_H$ the propagator reduces
to the black hole propagator, we proceed to compute the corrections when $%
r_s $ is close but not equal to $r_H$. It is easy to see that if the
dominant term is for example $(u-1)^\alpha$ then no corrections are obtained
from keeping terms of order $(u-1)^{\alpha+n}$ unless the term $%
(u-1)^{-\alpha}$ is included. The same occurs when $\mathrm{Im}\omega<0$ and 
$(u-1)^{-\alpha}$ dominates. Therefore, to compute the corrections we use
the approximations 
\begin{eqnarray}
\phi _{2}^{(\pm )} &\approx &\left\{A_{\pm } (u-1)^{\alpha }+B_{\pm
}(u-1)^{-\alpha }\right\} \\
f_{2}\partial _{r}\phi _{2}^{(\pm )} &\approx &2r\alpha \left\{ A_{\pm
}(u-1)^{\alpha }-B_{\pm }(u-1)^{-\alpha }\right\}
\end{eqnarray}
in the equation for the propagator.

With the above approximations, the expression for the correction factor $H$
becomes 
\begin{equation}
H(\omega ,0)\approx \frac{\lbrack \frac{i\omega }{r_{H}}-a](u-1)^{\alpha }-[%
\frac{i\omega }{r_{H}}+a]\frac{B_{-}}{A_{-}}(u-1)^{-\alpha }}{[\frac{i\omega 
}{r_{H}}-a](u-1)^{\alpha }-[\frac{i\omega }{r_{H}}+a]\frac{B_{+}}{A_{+}}%
(u-1)^{-\alpha }}\ ,
\end{equation}
where $a\equiv \partial_r\ln\phi_1$. With this expression it is very easy to
see that there are no poles in $G$. In the upper half plane, where the
imaginary part of $\omega $ is positive, $\alpha $ is negative and $%
G\rightarrow $ $-\frac{A_{-}}{A_{+}}$ as $r_{s}\rightarrow r_{H}$, ($%
u\rightarrow 1$). This expression has no poles where it is valid, i.e. in
the upper half plane. Similarly, in the lower half plane we have $%
G\rightarrow $ $-\frac{B_{-}}{B_{+}}$ which again has no poles where it is
valid. Let us now return to $H$. To find the zeroes and poles of $H$ , we
have to solve for the values of $\omega $ where the numerator or the
denominator vanishes.

We aim here to confirm what we saw numerically before, namely that the poles
and zeros accumulate on the real axis giving rise to a cut. In that case
what has to happen is that the poles go to zero as $r_s\rightarrow r_H$ (see
appendix for an example). To check this fact we only need to study the
propagator for frequencies such that $\omega/r_s<<1$ which is fortunate
since in that regime the equations simplify\footnote{%
If $\omega$ is small we are outside the validity of the quasistatic
approximation for the collapsing shell case, but here we want to check that
all poles converge towards zero. Moreover, the approximation is still valid
if we consider the other process which was discussed in the introduction,
the "slowed down" collapse at fixed ADM energy.} in two ways. First, the
mode in the interior of the shell, $\phi_1$ can be expanded as: 
\begin{equation}
\phi _{1}\approx \frac{1}{\sqrt{r}}\left\{ \frac{(\omega /2r)^{\nu }}{\Gamma
(\nu +1)}-\frac{(\omega /2r)^{\nu +2}}{\Gamma (\nu +2)}+\cdots -e^{i\pi \nu }%
\left[ \frac{(\omega /2r)^{-\nu }}{\Gamma (-\nu +1)}-\frac{(\omega
/2r)^{-\nu +2}}{\Gamma (-\nu +2)}+\cdots \right] \right\} \ .
\label{phi1app}
\end{equation}
Keeping only the leading term when $\omega/r_s\ll 1$ it follows that 
\begin{equation}
\phi _{1}\approx -e^{i\pi \nu }\frac{(\omega /2)^{-\nu }}{\Gamma (-\nu +1)}
r^{-\nu-\frac{1}{2}} \Rightarrow a=\partial_r \ln\phi_1 = \nu-\frac{1}{2}.
\end{equation}
Furthermore, we are able to do one more approximation: 
\begin{eqnarray}
\frac{B_{\pm }}{A_{\pm }} &=&\frac{\Gamma (2\alpha )\Gamma ^{2}(-\alpha
+\Delta _{\pm }/2)}{\Gamma (-2\alpha )\Gamma ^{2}(\alpha +\Delta _{\pm }/2)}
\nonumber \\
\mbox{} &\approx &-1+2i[\psi (\frac{\Delta _{\pm }}{2})+\gamma ]\frac{\omega 
}{r_{H}}+\cdots \approx -e^{-2i[\psi (\frac{\Delta _{\pm }}{2})+\gamma ]%
\frac{\omega }{r_{H}}}\ .
\end{eqnarray}
where the last exponential is introduced for convenience. Then, the two
equations (for poles and zeros) can be written in the form 
\begin{equation}
\left( \frac{a-ix}{a+ix}\right) e^{ib_{\pm }x}=1\ ,
\end{equation}
or equivalently 
\begin{equation}
\frac{x}{a}=\tan\left(\frac{xb_\pm}{2}\right) ,  \label{appeqn}
\end{equation}
with 
\begin{eqnarray}
x &=&\omega /r_{H}\ , \\
b_{\pm } &=&\ln (u-1)+2[\psi (\frac{\Delta _{\pm }}{2})+\gamma ]\ .
\end{eqnarray}
As can be easily seen by plotting the functions involved, the equation (\ref
{appeqn}) has an infinite number of real solutions which can be labeled by
an integer $n$. For $1\ll n \ll \ln((r_s^2-r_H^2)/r_s^2)$ at leading order
the solutions are 
\begin{equation}
\frac{\omega }{r_{H}}\approx -\frac{2\pi n}{\ln [\frac{r_{s}^{2}-r_{H}^{2}}{%
r_{H}^{2}}]+2[\psi (\frac{\Delta _{\pm }}{2})+\gamma ]}\ ,  \label{polezero}
\end{equation}
with $n$ an integer, corresponding to values on the real $\omega $ axis.
This is our result for the poles and zeroes of $H$ in the limit $%
r_{s}\rightarrow r_{H}$. If we choose $\Delta _{+}$ in (\ref{polezero}), we
obtain the poles, if $\Delta _{-}$, we obtain the zeroes. These two do not
agree as long as $\Delta _{+}\neq \Delta _{-}$, which is true for $\nu =%
\sqrt{1+m^{2}}>0$. Note that the solutions are purely real. We commented
earlier that as $r_{s}\rightarrow r_{H}$, the poles are complex, but flow
towards the real axis and the imaginary part quickly becomes many orders of
magnitude smaller than the real part. What happened above is that the
imaginary part was too small to show up without keeping more subleading
terms. In fact keeping the term of order $(\omega/r)^{\nu}$ in (\ref{phi1app}%
) gives rise to a tiny imaginary part, with a sign depending on the boundary
conditions used. In the extreme limit, the logarithm in the denominator of (%
\ref{polezero}) begins to dominate more and more over the $\Delta $%
-dependent term, so the poles and zeroes approach each other, finally
cancelling out and converting into a cut on the real axis. In the appendix
we give a simple example of how zeroes and poles can form a cut.

\section{Discussion}

In the previous sections we have shown how, as the shell moves closer to its
own horizon, the poles and zeros of the boundary propagator in the complex $%
\omega $ plane accumulate giving rise to a cut. The appearence of the cut is
the signal in the boundary theory of the formation of the black hole horizon
in the bulk. It is related to the fact that at the horizon different
boundary conditions (ingoing or outgoing waves) are appropriate on different
sides of the real $\omega $-axis. In our case these boundary conditions
appear automatically when the shell collapses. Once the cut has formed it is
meaningful to perform an analytic continuation of the propagator below the
real axis and then new poles can appear. In AdS$_{3}$ only the thermal poles
were found, but if we had performed the calculation in AdS$_{5}$ then the
poles computed in \cite{Horowitz:1999jd} would have appeared after the
collapse of the shell. Let us keep in mind that when we talk about collapse
and motion of the shell, it is only within the quasistatic approximation
which from the boundary perspective is more akin to a thermodynamic
treatment than a true dynamical description.

An open question is that there appears to be two relevant timescales as the
shell is approaching the horizon. On one hand, $\tau $, the inverse of the
imaginary part of a pole of $H$, gives a timescale corresponding to the
lifetime of the corresponding resonance. On the other hand, the equation of
motion for the shell (in this region) was approximately given by the
equation (\ref{freeze}) with a characteristic timescale 
\begin{equation}
\tau_H = 4\pi T_H \sim r_H \ .
\end{equation}
It remains to be understood properly which one is the relevant timescale for
the thermalization in this case: the lifetime of the resonances, or the time
for the collapse, and how the latter could be found in the boundary theory.

Our present understanding is then the following. In the beginning, as the
shell is very large, and just begins to collapse, the boundary theory is
very far from the thermal equilibrium. The propagator reveals this by having
a pole structure which is very different from a thermal propagator. The
poles correspond to an infinite tower of resonances in the boundary theory
with masses proportional to the shell radius, and lifetimes inversely
proportional to it. As the collapse begins to speed up we cannot apply the
quasistatic analysis - there will be dynamical processes in the boundary
theory which are beyond our control at this stage. However, as a first crude
step, we can jump to the very end of the collapse when a far-away observer
at a fixed radial distance would see the collapse to slow down and
``freeze'' as the shell is approaching the horizon radius. At that stage, as
a rough approximation, we again apply the quasistatic analysis, in the large
frequency domain. Now the boundary theory is approaching thermal equilibrium
with deviations encoded in the propagator. The poles and zeroes of the
propagator flow towards the real axis of the complex $\omega $-plane, and
finally degenerate into a cut with the propagator being the thermal one. The
full propagator is then defined either on\ the upper or lower half plane,
corresponding to a choice between a retarded or advanced propagator. A
continuation across the cut then reveals the thermal poles on the other
half-plane. I.e. $G=$ $-\frac{A_{-}}{A_{+}}$ is analytically continued down
through the cut into the lower half plane where one finds the thermal poles.
So the propagator has, as promised, become a thermal (retarded or advanced)
propagator when the shell is infinitesimally close to the horizon and
Hawking radiation begins to leak out and fill the anti-de Sitter space so
that a thermal equilibrium is established.

We emphasize that the propagator is just a measure of the properties of the
state of the boundary theory. The analysis does not tell us anything about 
\emph{how} the thermalization happens - there is no dynamical information
about the boundary theory. This question is tied up with going beyond the
quasistatic approximation, where dynamical issues become relevant. What one
would need at that stage, is a kinetic theory for the thermalization of the
strongly coupled boundary theory. This seems to be an untractable problem at
the moment. One can speculate that the equations governing the
thermalization of the boundary theory would be a kinetic theory equivalent
of the equations of motion of the collapsing shell in the bulk. That is, if
one were to understand the kinetic theory, one should be able to reproduce
the equation of the motion of the shell. Further, it has been suggested
before \cite{Susskind:1998vk, Polchinski:1999yd, Lowe:1999pk,
Susskind:1999ey} that a holographic description of even the simplest
dynamical processes in the bulk can be very non-local - even a slighest
innocuous looking approximation on the boundary side might completely
distort the bulk interpretation. The same could be true about the kinetic
theory / shell collapse duality. A small approximation to the kinetic
equations might completely mess up the bulk description of the collapse.

\bigskip

\section*{Acknowledgments}

We thank L. Thorlacius for a useful discussion. The work of U.D. was supported
by Swedish Natural Science Research Council (NFR) and that of M.K. by
The Swedish Foundation for International Cooperation in Research and Higher
Education (STINT).

\bigskip

\section*{Appendix}

In section 3 we discussed how the series of poles and zeros of the
propagator accumulate on the real axis giving rise to a cut. To clarify how
this happens we give in this appendix a simpler example in which this
occurs. Consider the function 
\begin{equation}
G(z)=\frac{\sqrt{\lambda }\Gamma (\lambda z)}{\Gamma (\lambda z+\frac{1}{2})}
\end{equation}
where $\lambda $ is a positive real number ans $z\in C$. The function $G(z)$
has poles at $z=-n/\lambda $ and zeros at $z=-(n+1/2)/\lambda $ with $n$ a
positive integer. When $\lambda \rightarrow \infty $ the poles and zeros
accumulate on the negative real axis. In that limit and for $-\pi <\arg
(z)<\pi $ the Stirling approximation gives 
\begin{equation}
\lim_{\lambda \rightarrow \infty }G(z)=\lim_{\lambda \rightarrow \infty }%
\frac{\sqrt{\lambda }\Gamma (\lambda z)}{\Gamma (\lambda z+\frac{1}{2})}%
=\lim_{\lambda \rightarrow \infty }\sqrt{\lambda }\frac{(\lambda z)^{\lambda
z-1/2}e^{-\lambda z}}{(\lambda z+1/2)^{\lambda z}e^{-\lambda z-1/2}}=\frac{1%
}{\sqrt{z}}
\end{equation}
The resulting function, $1/\sqrt{z}$ has a cut which, as follows from the
calculation, is placed on the negative real axis. In fact this example
arises when taking the large frequency limit of the AdS propagator in global
coordinates. In that case one recovers the propagator in Poincar\'{e}
coordinates which has a cut on the positive real axis.

\end{document}